\RequirePackage{fix-cm}
\documentclass[twocolumn]{svjour3}          
\smartqed  
\usepackage[pdftex]{graphicx}

\graphicspath{{figures/}}

\usepackage{url}
\usepackage[colorlinks=true,citecolor={blue},linkcolor={blue},urlcolor={blue}]{hyperref}
\usepackage[numbers,round,authoryear]{natbib}

\usepackage{amsmath,amssymb,amsfonts,amsthm}

\usepackage{siunitx}
\usepackage{enumerate}
\usepackage{enumitem}

\usepackage{subfig}
\usepackage{caption}

\begin{document}

\newcommand{\D}{\mathbb{D}}
\newcommand{\R}{\mathbb{R}}
\newcommand{\Q}{\mathbb{Q}}
\newcommand{\Z}{\mathbb{Z}}
\newcommand{\EE}{\mathbb{E}}
\newcommand{\PP}{\mathbb{P}}
\newcommand{\dd}{\textrm{d}}

\title{An integrate-and-fire model to generate spike trains with long-range dependence
}


\author{Alexandre Richard         \and
        Patricio Orio \and Etienne Tanr\'e
}


\institute{A. Richard \at
              CentraleSup\'elec, Universit\'e Paris-Saclay, Laboratoire MICS et F\'ed\'eration CNRS - FR3487, France. \\
              \email{alexandre.richard@centralesupelec.fr}           
           \and
           P. Orio \at
              Instituto de Neurociencia, Facultad de Ciencias, Universidad de Valpara\'iso and Centro Interdisciplinario de Neurociencia de Valpara\'iso, Universidad de Valpara\'iso, Chile.\\
		\email{patricio.orio@uv.cl}
	\and
		E. Tanr\'e \at
		Universit\'e C\^ote d'Azur, Inria, 2004 Route des Lucioles BP 93, 06902 Sophia-Antipolis, France.\\
		\email{Etienne.Tanre@inria.fr}
}

\date{Received: date / Accepted: date}

\maketitle

\begin{abstract}
Long-range dependence (LRD) has been observed in a variety of phenomena in nature, and for several years also in the spiking activity of neurons. Often, this is interpreted as originating from a non-Markovian system. Here we show that a purely Markovian integrate-and-fire (IF) model, with a noisy slow adaptation term, can generate interspike intervals (ISIs) that appear as having LRD. However a proper analysis shows that this is not the case asymptotically. For comparison, we also consider a new model of individual IF neuron with fractional (non-Markovian) noise. The correlations of its spike trains are studied and proven to have LRD, unlike classical IF models. On the other hand, to correctly measure long-range dependence, it is usually necessary to know if the data are stationary. Thus, a methodology to evaluate stationarity of the ISIs is presented and applied to the various IF models. We explain that Markovian IF models may seem to have LRD because of non-stationarities.
\end{abstract}

\section{Introduction}
\label{intro}

The modelling of neuronal activity has a long and rich history whose first successes date back to the 50's and the seminal work of \cite{HH}. A few years later, a simpler probabilistic model based on the passage times of a random walk was introduced by \cite{GersteinMandelbrot}, corresponding to a stochastic version of the Perfect Integrate-and-Fire (PIF) model.

The activity of a neuron is characterised by the electrical potential of its membrane, and more precisely by spikes whose amplitude and duration are very similar to one another. Therefore, it is rather the sequence of times at which these spikes occur which is believed to carry the neuronal information. While temporal (and spatial) correlations between interspike intervals (ISIs) have been observed for a long time (see \citep{ChacronEtAl2003} and references therein), the presence of fractal behavior \citep{Teich92,BairEtAl94} and LRD phenomena in the spiking activity of neurons has been acknowledged for only two decades: see \citep{TeichEtAl96,TeichEtAl97,LewisEtAl01,LowenEtAl01,BhattacharyaEtAl}, including artificially grown neuronal networks in \citep{SegevEtAl02}, etc. (see the introduction of \citep{Jackson} for a very comprehensive list of references). This LRD phenomenon is ubiquitous in nature, and takes the form of power-law correlations between interspike intervals rather than exponentially decaying correlations. In particular, LRD implies that the present neuronal activity is correlated with a very distant past. 

Until recently in the neuroscience literature, long-range dependence, also called long memory, has been quantified mostly by the Fano factor. In \citep{BhattacharyaEtAl}, temporal and spatial LRD of \emph{in vivo} human hippocampal neurons is detected relying on statistics like the detrended fluctuation analysis \citep{PengEtAl92}. We shall adopt a similar approach, which has also been used to detect LRD in ion channels \citep{CorrelChannels}. LRD may arise due to the influence of presynaptic neurons, as well as intrinsic factors such as fluctuations in ion channel activity (producing LRD in neurotransmitter exocytosis, as described by \cite{LowenEtAl97}). \cite{SchwalgerEtAl2015} also mention several possible sources of LRD: neural refractoriness, bursting and adaptation. Here, the LRD phenomenon is closely related (although not rigorously equivalent) to the power-law decay of correlations of the ISIs. The latter property has been considered as a near optimal way of encoding neuronal information (\cite{BaddeleyEtAl}).

Early attempts to replicate the LRD property of ISIs were based on point processes models and were proposed by \cite{Teich92,BairEtAl94}, and more recently by \cite{Jackson}. Instead, we focus here on stochastic Integrate-and-Fire models, especially because they allow to preserve the aforementioned interpretation on the origin of LRD. Besides, it is commonly accepted that they provide a good compromise between biologically complex and realistic models such as the Hodgkin-Huxley model, and more simple and amenable ones to perform statistical computations with. 

\cite{BrunelSergi} and \cite{DestexheEtAl03} noticed that an additional differential equation for the synaptic current, coupled with the membrane potential equation of a simple IF model, introduces temporal correlations in the dynamics. Assuming that the pre-synaptic excitation is modelled by a Poisson noise, it is natural by diffusion approximation to write the synaptic equation as a stochastic differential equation driven by white noise. An interesting feature of this model is that it is simple enough to compute (or approximate) some ISI statistics: for example, \cite{MiddletonEtAl2003} focused on the ISI density, power spectral density and Fano factor of the PIF, \cite{Lindner2004} on serial correlation coefficients of the PIF, \cite{SchwalgerEtAl08}  on the ISI density, coefficient of variation, Fano factor of the leaky integrate-and-fire (LIF) model, etc. We also refer to \citep{SacerdoteGiraudo} for a mathematical and statistical treatment of Markovian IF models.

The purpose of this paper is to explain that a (linear) IF model with Markovian noise, even enhanced with a noisy adaptation variable, has exponentially decaying correlations which cannot produce long-range dependent ISIs. To account for different correlation patterns observed on real data, we introduce an IF model governed by a non-Markovian noise, namely a fractional Brownian noise. The fractional Brownian motion (fBm) is a stochastic process whose increments (the noise process) are long-range dependent and stationary. It naturally appears in modelling as a limit of more simple processes: For instance, the fBm appears as the limit of high-dimensional Orstein-Uhlenbeck processes \citep{CarmonaCoutinMontseny}. However, it is non-Markovian, which makes it a challenge to study and compute all the aforementioned statistics of spike trains. We shall discuss the related idea developed by \cite{SchwalgerEtAl2015}, where general Gaussian processes are proxied by finite-dimensional Markov processes and serve as input in an IF model.

In addition to modelling, our contribution is also methodological: we compare several measures of LRD and stationarity. Indeed, testing stationarity is important in the attempt to measure LRD, as we shall see that non-stationary spike trains from Markovian models can give the illusion of LRD. We refer to \citep{Samorodnitsky} and \citep{Beran} on these questions, as well as the collection of review articles edited by \cite{RangarajanDing} on modelling long-range dependent phenomena in various fields ranging from economy, biology, neuroscience to internet traffic. Last but not least, one is often interested in getting estimates on the distribution of the ISIs. But without the stationarity assumption, these distributions are likely to vary with time, which makes the estimation procedure either difficult or inaccurate. Hence, it is crucial to determine if these distributions vary with time, as it is desirable that the sequence of ISIs be in a stationary regime for such study. We therefore explain how to test this assumption, with a direct application to ISIs generated by integrate-and-fire models.

The remainder of the paper is organized as follows: in Section~\ref{sec:stats}, we present an account of the tools and methods to measure LRD and stationarity from a single spike train. Then the stochastic Integrate-and-Fire models and some of its variations are presented in Section~\ref{sec:models}, with an emphasis on fractional noise. The results of our analysis are detailed in Section~\ref{subsec:PIF} for the PIF with Markovian noise with or without adaptation, in Section~\ref{subsec:resultsfPIF} for the PIF with fractional noise, and in Section~\ref{subsec:otherPIFS}  for variants with mixed Brownian and fractional noise. Finally, we discuss these results and compare them to previous models in Section~\ref{sec:discussion}.

\section{Methods: Statistical measurement of long-range dependence and stationarity}\label{sec:stats}

\subsection{Long-range dependence}\label{subsec:LRD}

The terminology ``long memory'' or ``long-range dependence'' appeared in the early work of Mandelbrot and coauthors in the 60's \citep{Mandelbrot65,MandelbrotWallis68}, in an attempt to describe the phenomenon observed by Hurst on the flows of the Nile river.

If $X$ is a random variable or a stochastic process, we say that $X(\omega)$ is a realization (or an observation) of $X$ for the outcome $\omega$ in the probability space $\Omega$ of all possible outcomes. Let us denote by $\EE$ the expectation of a random variable. A 
sequence of random variables $\{X_n\}_{n\in\mathbb{N}}$ has the \emph{long-range dependence (LRD)} property if it satisfies:
\[
\sum_{n=1}^{\infty} \EE\left[(X_1 - \mathbb{E}X_1)( X_n - \mathbb{E}X_n)\right] = +\infty .
\]
Observe that the LRD property is obtained by averaging over all possible outcomes. In practical situations though, where we might have access to very few realizations (or even a single one) of the same phenomenon, at least two limitations appear: the length of the sequence is finite, and we do not know the law of the $X_n$'s (in fact when dealing with spike trains, 
we often have only one sample of the sequence). To detect long-range dependence, 
we will use two estimators: the detrended fluctuation 
analysis (DFA) and the rescaled range statistics ($R/S$). There exist other popular methods to measure the Hurst parameter (properly defined in Section~\ref{subsec:rrs}), but as seen from \citep{Taqqu,Weron} we may not expect to get much better results than with the DFA and $R/S$ methods.
Besides, the latter is the only 
statistics for which it has been possible to prove convergence to the 
Hurst parameter rigorously in some non-trivial cases as the number of observations goes to infinity \citep{Samorodnitsky,Beran}.\\
To prove convergence of the $R/S$ statistics, it is usually required 
that the sequence $\{X_n\}_{n\in\mathbb{N}}$ is $L^2$-stationary, in the sense that
\begin{equation}\label{eq:L2stat}
\begin{split}
&\text{for all } n,\quad \EE(X_n) = \EE(X_1) , \\ 
&\text{for all } n\geq m,\quad \EE(X_n X_m) = \EE(X_{n-m+1} X_1) ,
\end{split}
\end{equation}
although there are examples of such convergence for non-stationary data (\citep{Bhattacharya83} and \citep[p.183--185]{Samorodnitsky}). Verifying this requirement is often eluded in practical situations, although non-stationarity may have important consequences on the interpretation of statistical analysis of data. We emphasize that measuring (non-)stationarity and long-range dependence is a tricky question.

Let us insist on the type of data we shall be dealing with: these are (finite) 
sequences $X_1, \dots, X_N$ (we now use this notation both for the probabilistic 
model and a realization of it). We aim at obtaining the Hurst parameter of the data from a single sequence (i.e. not from averaging several realizations), to cope with biological constraints.

\subsubsection{The rescaled ranged statistics (\protect{$R/S$})}\label{subsec:rrs}

For a sequence $\{X_n\}_{n\in \mathbb{N}}$ of random variables, let $\{Y_j = \sum_{i=1}^j X_i\}_{j\in \mathbb{N}}$ denote the sequence of the cumulated sums, and let the rescaled-range statistics be defined as:
\[
R/S(N) = \frac{\max_{1\leq i\leq N}(Y_i - \frac{i}{N} Y_N) - \min_{1\leq i\leq N}(Y_i - \frac{i}{N} Y_N)}{\sqrt{\frac{1}{N}\sum_{i=1}^N (X_i- \frac{1}{N} Y_N)^2}} \ .
\]
If for some $H\in (0,1)$, the law of $\frac{1}{N^H} R/S(N)$ converges, as $N$ goes to $+\infty$, towards some positive random 
variable denoted by $e^\mathbf{b}$, we call $H$ the Hurst parameter of the 
model. In the most simple example, where the $X_n$'s are independent and 
identically distributed (\emph{i.i.d.}) with finite variance, the convergence 
occurs with $H=0.5$. We consider that data have LRD when 
$H>0.5$ (the reverse case $H<0.5$ is often called 
anti-persistence, but we will not encounter it here).

Let us recall that $N$ denotes the length of the sequence of data 
$X_1, \dots, X_N$. A simple way to estimate $H$ 
is to fit the following linear model for various values of $N$:
\begin{equation*}
\log R/S(N) = \mathbf{b} + H\ \log N \ .
\end{equation*}
However this is not the robust way to proceed in practice, see 
\citep{Beran,Taqqu,Weron}. Instead, we divide the data into $M$ blocks of length $n$ ($N = M\times n$) and compute the $R/S$ statistics on each block 
$\widetilde{R/S}(m,n)$ for $m=1 \dots M$. Then, we average over all blocks 
to obtain $\widetilde{R/S}(n) = \tfrac{1}{M} \sum_{m=1}^M 
\widetilde{R/S}(m,n)$. Finally we let $n$ take integer values between $1$ and $N$ and estimate the slope of the function $n\mapsto \widetilde{R/S}(n)$. This slope gives the estimated Hurst parameter of $\{X_i\}_{i\leq N}$, frequently denoted by $\hat{H}_N$ in the rest of this paper.

Let us conclude this paragraph with several insightful examples:
\begin{itemize}[topsep=0pt,itemsep=0pt,partopsep=1ex,parsep=1ex,leftmargin=*]
\item If the $X_n$'s are \emph{i.i.d.} and $\EE\left(X_n^2\right)<\infty$, then standard convergence results imply that $\frac{{1}}{\sqrt{N}} R/S(N)$ converges.
\item If the $X_n$'s are mixing and stationary, then $H=0.5$ (see Section \ref{subsec:heuristics}).
\item If the $X_n$'s are the increments of a fractional Brownian motion with scaling parameter $\alpha$ (\emph{fBm}, see Section \ref{subsec:noise}), then $\alpha$ is also the Hurst parameter, i.e. $\frac{1}{N^\alpha} R/S(N)$ converges.
\end{itemize}

There are also examples of sequences of random variables with infinite variance such that $\frac{{1}}{\sqrt{N}} R/S(N)$ converges \citep[p.178--180]{Samorodnitsky}, which emphasizes the robustness of the $R/S$ method to a wide class of distribution of the $X_n$. There are also examples of non-stationary random sequences for which the $R/S$ statistics converges at prescribed rate $H\in [0.5,1)$ \citep[p.187]{Samorodnitsky}.

\subsubsection{The detrended fluctuation analysis (DFA)}

This method was introduced by \citet{PengEtAl92,PengEtAl94} in genetics. We merely rephrase \citep{Weron} to present it. See also \citep{Taqqu} where it is called Residuals of Regression method and where it is compared to other methods.

Like in the $R/S$ analysis, the data are divided into $M$ blocks of length $n$. For $m=1\dots M$ and $j=1\dots n$, we denote the partial sum on block $m$ by $$Y_{m,j} = \sum_{i=1}^j X_{(m-1)n + i}.$$ On each block, a linear regression is applied to determine coefficients $(a_m,b_m)$ such that $\tilde{Y}_{m,j}:= a_m j + b_m, j=1\dots n$ is the least-square approximation of $Y_{m,j}$. Then, the empirical standard deviation of the error is computed:
\begin{equation*}
s_m := \sqrt{\frac{1}{n} \sum_{j=1}^n \left(Y_{m,j} - \tilde{Y}_{m,j} \right)^2 } .
\end{equation*}
Finally, the mean of these standard deviations is $\bar{s}_n := \tfrac{1}{M} \sum_{m=1}^M s_m$. The analysis performed with $\widetilde{R/S}(n)$ can now be reproduced with $\bar{s}_n$ (i.e. the heuristics is that $\bar{s}_n$ behaves asymptotically as a constant times $n^H$). The slope computed from the log-log plot is again denoted by $\hat{H}_N$.

\subsubsection{Surrogate data} 
To check for the statistical significance of the $R/S$ and DFA analyses, we employed a bootstrapping with replacement procedure. For each simulated spike train, we produced $100$ sequences of spikes by randomly shuffling  the interspike intervals of the spike train. In this way, we obtain $100$ new spike trains having the same interspike interval distribution, but without any correlation structure between spikes. The LRD analysis is applied to these new data to estimate $\hat{H}_N$ as a function of $N$ for each of them. The mean over all surrogate samples is plotted (see Section~\ref{sec:results}) in a black solid line, while two gray lines represent the mean $\pm$ 2 times the standard deviation. Thus, the region between gray lines will contain roughly 95\% of possible $H$ values that can be obtained by chance from a non-correlated data series. If the plot of $\hat{H}_N$ of the initial spike train enters this shadow region, then it is doubtful that the data have the LRD property.

\subsection{Stationarity}

To compute statistics on a time series (e.g. spike trains) such as the mean, the distribution or more complex statistics aimed at determining the presence of power law or LRD, it is often necessary that the series is stationary. But it is in general a difficult problem to decide whether time series data are issued from a stationary distribution or not. Like the measurement of long-range dependence, part of the difficulty here arises from the fact that we want to decide whether biological data are stationary relying on a single observation (i.e. a single sequence of spikes).

Here we used several tests for stationarity: a simple windowed Kolmogorov-Smirnov (KS) test, the Priestley-Subba Rao (PSR) test and a wavelet-based test.  
Note that notion of stationarity itself must be clarified: the first test (KS) evaluates \emph{strong stationarity}, i.e. whether the law of the process is invariant by any time shift. The PSR and wavelet-based tests consider a weaker form of stationarity that we shall refer to as \emph{$L^2$-stationarity}. A process $X$ is $L^2$-stationary if it satisfies Equation (\ref{eq:L2stat}). It is important to have in mind that the best test to use in a given situation depends strongly on the type of non-stationarity of the data (see for instance Table 2 in \citep{CardinaliNason10}). Since we do not know \emph{a priori} what type of non-stationarity may appear, we are applying several tests.\\
Note that a frequently used test for stationarity is the KPSS test \citep{KPSS}, based on unit root testing. However, we found unit root tests to perform badly when used on fractional noise (which is stationary). 

Like for the Hurst estimation of the previous section, these tests are designed to be run on a single realization of the process (said otherwise, no averaging is needed), which is well-suited for biological data. However, to decide whether a model yields stationary spike trains, multiple simulations can be performed. Hence the PSR and wavelet-based tests were applied to $50$ simulations of the same model and a boxplot of the p-values was plotted. If the data come from a model which produces stationary ISIs, p-values must be uniformly distributed between $0$ and $1$. Otherwise we may conclude that the model yields data which are not stationary.

Thus, this methodology is designed to decide whether our models produce stationary ISIs. If the problem is to decide whether a single (biological or simulated) sequence is stationary, such stationarity tests will merely give a probability that the sequence is stationary.

\subsubsection{The windowed Kolmogorov-Smirnov (KS) test}\label{subsubsec:windowKS}

Based on the usual Kolmogorov-Smirnov (\emph{KS}) test, we designed a windowed KS test. In this test, the ISI series are split in windows of fixed time length. Each block is tested against the others to see if they are described by the same distribution, using the non-parametric KS test. For each pair, the p-value is then represented in a two-dimensional table. The null hypothesis is that the two samples are drawn from the same distribution. Hence, small p-values indicate that the data may be non-stationary. In this way, a visual map is obtained in which one can easily detect portions of the time series that do not follow the same distribution as the others. Since this test does not return a single p-value, it is not suited to the aforementioned methodology of repeating simulations. Yet it allows for simple interpretations and we keep it for comparison with other tests.

\subsubsection{The Priestley-Subba Rao (PSR) test}

Let $X = \{X_t\}_{t\in \R}$ be a centered stochastic process with finite variance. It is known that if $X$ is $L^2$-stationary, then
\begin{equation*}
X_t(\omega) = \int_\R e^{i f t} A(f)\  Z_\omega(\dd f) ,
\end{equation*}
where $Z = \{Z_\omega(B),\ \omega\in \Omega, B\in \mathcal{B}(\R) \}$ is a random measure on $\R$ and $A$ is the spectral density function. A natural generalization of the definition of $X_t$ is to let $A$ depend on time. The PSR test \citep{PSR} evaluates the time dependence of $A_t(\cdot)$. Thus, a test is proposed with the following null hypothesis: $t\mapsto A_t(f)$ is constant in time.
Note that the process must have zero mean (data can be centered in practice), finite variance, and be ``almost'' Gaussian.

A two-factor analysis of variance is performed. If the first p-value is small, then the test can stop here and the data declared to be non-stationary \citep{PSR}. Otherwise, one can proceed to test the stationarity with a second p-value.

\subsubsection{A wavelet-based test}

This test \citep{CardinaliNason16} is designed for a large class of processes called locally stationary wavelet processes, which can be written:
\begin{equation*}
X_t(\omega) = \sum_{j=1}^\infty \sum_{k=-\infty}^\infty \theta_{j,k} \psi_{j,k}(t) \xi_{j,k}(\omega) ,
\end{equation*}
where $\{\psi_{j,k}(t), t\in \Z \}$ is a wavelet basis, $\{\xi_{j,k}\}$ is an array of \emph{i.i.d.} random variables with mean $0$ and variance $1$, and $\{\theta_{j,k}\}$ are the (deterministic) wavelet coefficients of $X$.\\
A test statistic is constructed from the data and a p-value is computed to decide whether the so-called $\beta$-spectrum (see \citep{CardinaliNason16}) is constant. If it is, the data are then stationary.
We refer to \citep{Nason,CardinaliNason16} and references therein for more details about this test.

\vspace{0.3cm}

The PSR test and this wavelet test give excellent results when applied to
pure fractional noise, in the sense that they repeatedly give large p-values, as expected.

\subsection{Numerical tools}

To test stationarity, we relied upon \emph{Python}'s function \texttt{stats.ks\_2samp} from the \texttt{scipy} library for our windowed KS test, and upon the couple of \emph{R} packages:
\begin{itemize}[topsep=3pt,itemsep=0pt,partopsep=0pt,parsep=1pt]
\item for the PSR test, we have used the \emph{R} package \texttt{fractal}, and particularly the function \texttt{stationarity}.
\item for the wavelet-based test, we have used the function \texttt{BootWPTOS} from the \emph{R} package \texttt{BootWPTOS}.  
\end{itemize}
Our \emph{Python} code to measure the Hurst parameter and to generate spike trains from the various models presented hereafter is available in modelDB \\
(http://modeldb.yale.edu/235054).

\section{The models}\label{sec:models}

We describe a large class of noisy integrate-and-fire models 
with adaptation. Integrate-and-fire models have two regimes.
The \emph{subthreshold regime} is characterized by the stochastic differential system
\begin{align}
\begin{split}
\dd V_t &= \left(\mu_V - \lambda_V V_t + \gamma Z_t\right)\dd t + \sigma \dd B^\alpha_t \\
\dd Z_t &= \left(\mu_{Z,t} - \lambda_Z Z_t\right)\dd t + \sigma' \dd \tilde{B}^\alpha_t.
\end{split}\label{eq:generalLIF}
\end{align}
The process \((V_t,t\geq 0)\) models the membrane potential (normalized between 0 and 1) and \((Z_t, t\geq 0)\) corresponds to an adaptation variable. We call $Z$ the adaptation variable/process, even though in several cases we remove the adaptation mechanism. \(\mu_V\), \(\lambda_V\), \(\lambda_Z\), \(\gamma\), \(\sigma\), \(\sigma'\) and $\alpha$ are parameters of the model. We detail the role of $\mu_{Z,t}$ in the next paragraph. \(\mu_V\) is the voltage offset (in $ms^{-1}$); \(\lambda_V\) is the relaxation rate of the voltage (in $ms^{-1}$); \(\gamma\) is the coupling factor between the adaptation variable \(Z_t\) and \(V_t\) (in $ms^{-1}$); \(\lambda_Z\) is the relaxation rate of the adaptation (in $ms^{-1}$); \(\sigma\) and \(\sigma'\) are the intensities of the noises $B^\alpha$ and $\tilde{B}^\alpha$ (in $ms^{-\alpha}$) -- random noises called fractional Brownian motions and described further in Paragraph \ref{subsec:noise}.

\(\mu_{Z,t}\) is an offset factor for \(Z\). We will either consider that \(\mu_{Z,t}\) is constant in time ($\mu_{Z,t}\equiv \mu_Z$) or that it varies during 1 ms only after a spike ($\mu_{Z,t}\equiv \mu_Z + \epsilon_{Z,t}$). In both cases, let us remark that the law of \((V_t,t\geq 0)\) remains invariant by the modification of parameters \((\mu_Z,\mu_V)\rightarrow (\mu_Z + a, \mu_V - \gamma \frac{a}{\lambda_Z})\). So, to reduce the number of parameters to estimate, we assume that \(\mu_Z = 0\). In the second case (adaptation), we thus have $\mu_{Z,t}= \epsilon_{Z,t}$, where $\epsilon_{Z,t}$ is equal to 1 during 1 ms after a spike and $0$ otherwise. Using this form of adaptation (instead of a fixed increment of $Z$) puts a natural limit to $Z$, mimicking the behavior of a finite population of ion channels \citep{Schwalger}. More hidden states like $Z$ can be added in (\ref{eq:generalLIF}) to approximate Gaussian processes which have long-range correlations (see \citep{SchwalgerEtAl2015}) where this idea is fully developed, and Subsection~\ref{subsec:HD_IF} where we test it numerically.

The \emph{firing regime} is activated at the times \(\tau\) when the membrane potential hits
a fixed (deterministic) threshold \(V^{\mathrm{th}}\).
We call such a time \(\tau\) a firing time.
Just after \(\tau\), the membrane potential is \textit{reset} to 
a fixed value \(V^{\mathrm{r}}\), the rest potential. At the same time, we recall that $\mu_{Z,t}$ can be incremented due to adaptation. The sequence of firing times is formally defined for $n=1,2,\dots$ as
\begin{equation*}
\tau_n = \inf\{t\geq \tau_{n-1}: V_t = V^{\mathrm{th}} \}
\end{equation*}
and $\tau_0 = 0$. The sequence of interspike intervals is $\{X_n = \tau_n-\tau_{n-1}\}_{n\in \mathbb{N}}$, consistently with the notations of Section~\ref{sec:stats}.

\subsection{The noise}\label{subsec:noise}
In Equation (\ref{eq:generalLIF}), the noises \((B_t^\alpha,t\geq 0)\) and  \((\tilde{B}_t^\alpha,t\geq 0)\) are fractional Brownian motions (fBm) of scaling parameter \(\alpha\in (0,1)\). This family of processes, introduced in \citep{MandelbrotVN}, is given by Gaussian centered processes with covariance function
\[
\mathbb{E}\left[B^\alpha_t B^\alpha_s\right] = \frac{1}{2}
\left(|t|^{2\alpha}+|s|^{2\alpha}-|t-s|^{2\alpha}\right).
\]
The case \(\alpha=0.5\) corresponds to the standard Brownian motion (integral of white noise). When $\alpha>0.5$, the increments of the fBm (i.e. the fractional noise) have positive correlations decaying very slowly, according to a power law:
\begin{align*}
\EE\left[B^\alpha_1 (B^\alpha_{n+1}-B^\alpha_n)\right] &= \frac{1}{2} \left((n+1)^{2\alpha} + (n-1)^{2\alpha} - 2n^{2\alpha}\right) \\
&\sim 2\alpha(2\alpha-1) n^{2\alpha-2} .
\end{align*}
This property will account either for a strongly correlated synaptic input to the neuron, or it could also be that the membrane acts as a fractional integrator over a noncorrelated input noise. Contrary to most noises encountered in the literature (in particular Markovian noises), the range of dependence of this noise can be said to be infinite. Mathematically, this is the long-range dependence property we shall include in our models. A complementary interpretation of $\alpha$ is as a scaling parameter (or power law): indeed, the fBm is statistically scale invariant of parameter $\alpha$, meaning that an homothetic time change of parameter $c$ will result in an homothetic space change of order $c^\alpha$.

This stochastic process has already been applied in various fields of physics and more recently, biology: in the context of biological dynamics (cell movement in crowded environment, so-called anomalous diffusions), see for instance \citep{MetzlerKlafterReview,ChurillaEtAl,RangarajanDing}. Notably, a reviewer kindly pointed to us the work of \citep{ZilanyEtAl} that makes use of fractional Brownian motion in the modelling of auditory-nerve fibers: As a consequence, their model displays power law adaptation properties, as biologically observed several years earlier by \citep{FairhallEtAl01}.\\
More generally, fractional Brownian motion provides a good, mathematically tractable, model of so-called $1/f$ noise, see e.g. \citep{AbryEtAl95}. $1/f$ noise has been successfully applied to describe many phenomena, from heartbeat \citep{Peng93LRD} to internet traffic \citep{Willinger97}, including neuronal fractal dynamics \citep{LowenEtAl01,SobieEtAl}.

Contrary to standard Brownian motion, the fBm with $\alpha\neq 0.5$ is not Markovian, which makes the computation of even basic statistics of ISIs a very difficult problem. The case $\alpha<0.5$ also yields power-law correlations, but negative, which is not used here. Very little is known on the first-passage time (i.e. spiking time) of models such as (\ref{eq:generalLIF}) driven by fractional Brownian motion: for the passage-time of fBm itself, see \citep{DelormeWiese} for simulations and formal estimation of the density when $\alpha$ is close to $0.5$ and \citep{DecreusefondNualart} for inequalities on its Laplace transform, and on the general model see \citep{RichardTalay} for inequalities on Laplace transforms.

\subsection{Models without the $Z$ variable \protect{($\gamma = 0$)}}
When $\gamma=0$, (\ref{eq:generalLIF}) is a noisy Leaky Integrate-and-Fire (LIF) model. The particular case \(\lambda_V=0\) corresponds to the noisy Perfect Integrate-and-Fire (PIF) model. The membrane potential 
is solution of a linear stochastic differential equation. 
In the white noise setting \(\alpha=0.5\),
the interspike intervals are independent and identically distributed, so in particular such sequences are stationary and do not have LRD.

Compared with multidimensional Markov models (see \ref{subsec:HD_IF}), this model with $\alpha>0.5$ is also more compact. This can be interesting when one needs to estimate the parameters from real data, see \citep{SacerdoteGiraudo}.

We have also chosen to consider a model without any refractory period. The results seem to be interesting even in this simplified case.

\subsection{Simulation tools}

We simulated the subthreshold regime (\ref{eq:generalLIF}) with a simple Euler scheme. The hitting times are then recorded each time the simulated value of $V$ reaches a value above the threshold, according to the firing regime described above. Note that there is no simple and efficient algorithm to simulate fractional Brownian motion. For our simulations, we chose the most efficient \emph{exact} algorithm, namely the Davis-Harte algorithm \citep{Coeurjolly} (this algorithm has complexity $O(n\log n)$). 
All our Python code is available in \emph{modelDB} and can be downloaded from http://modeldb.yale.edu/235054.

\section{Study of LRD and Stationarity of the simulated data}\label{sec:results}

\subsection{PIF model with stochastic adaptation ($\alpha=0.5$)}\label{subsec:PIF}

\subsubsection{Long-range dependence}\label{subsec:PIF_LRD}

We simulated a spike train using a Perfect Integrate-and-fire (PIF) model with adaptation, i.e. \eqref{eq:generalLIF} with $\lambda_V=0$. The voltage dynamic is deterministic ($\sigma=0$) and the slow adaptation variable ($Z$) has an additive white noise (i.e. $\alpha=0.5$)  with $\sigma'=\num{2.5e-6}$.

\begin{figure}[!th]
\centering
\includegraphics[width=0.9\linewidth]{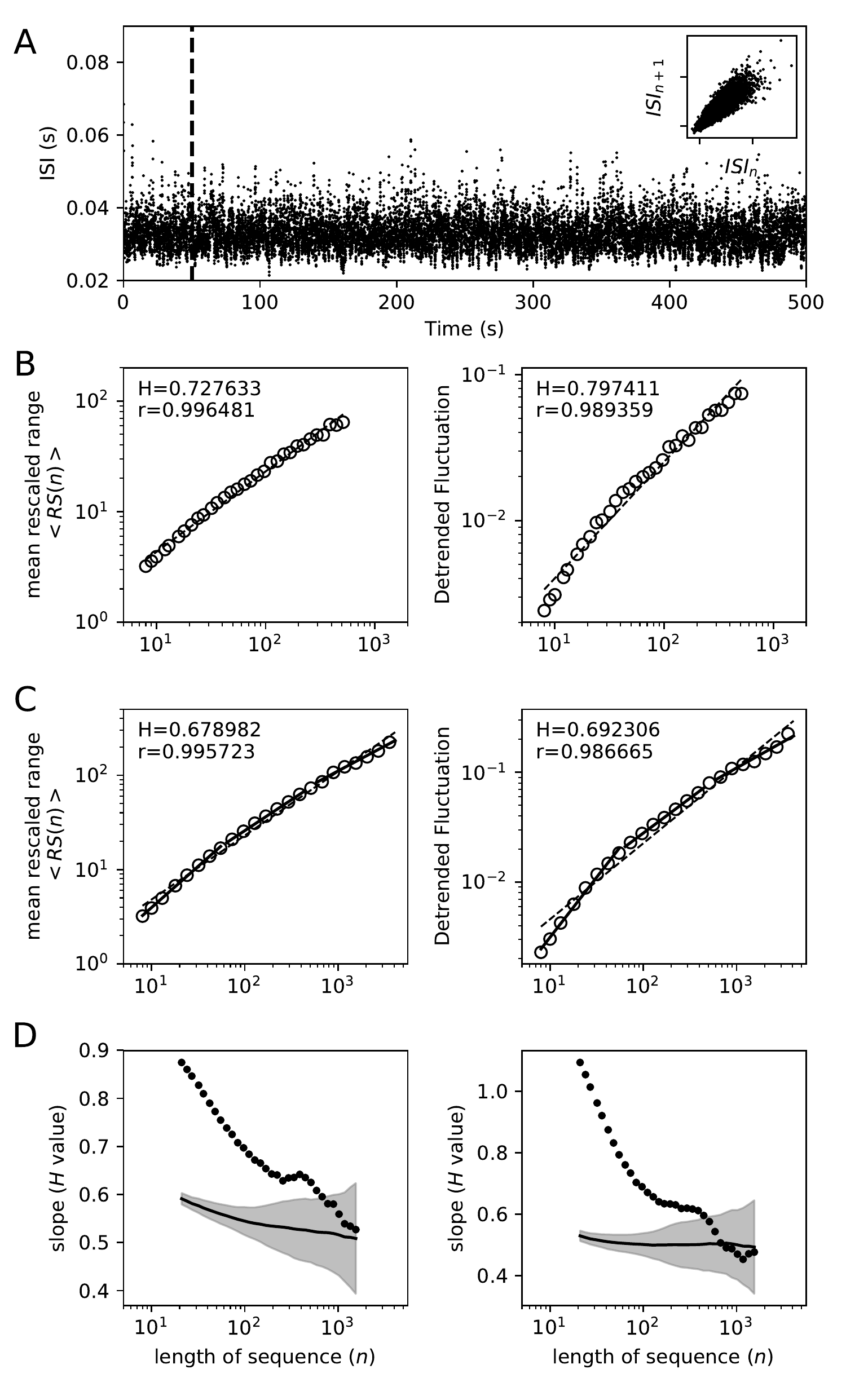}
\caption{$R/S$ and DFA analysis of PIF model with noisy adaptation. \textbf{A.} ISI sequence analyzed. parameters are $\mu_V=0.04$, $\lambda_V=0$, $\gamma=-0.3$, $\sigma=0$, $\lambda_Z=0.005$, $\sigma'=\num{2.5e-6}$  (all in units of $[ms^{-1}]$). The vertical segmented line shows the limit of the data analyzed in \textbf{B}. Inset, $ISI_{n}/ISI_{n+1}$ plot. \textbf{B.} Rescaled range (left) and Detrended Fluctuation Analysis (right) for the ISIs in the first 100 seconds of simulation (3023 spikes). H value is the slope of the best fit of $\log n$ vs. $\log R/S(n)$ or $\log DFA(n)$ points to a straight line (segmented line over the data points). \textbf{C.} $R/S$ and DFA analysis of the full ISI sequence (15164 spikes). 
The $H$ values indicated in the top left corner, and the segmented lines correspond to the fit of the full set of points to the data as in \textbf{B}. The shorter, continuous lines depict the best fit of a subset of the points. For clarity, not all points are shown. \textbf{D.} Slope values calculated at different $n$ values (with a moving window of 15 points), for the $R/S$ (left) and DFA analysis (right). The continuous line shows the mean slope calculated with 100 surrogate series and the shadow region shows the empirical Standard Deviation of the surrogate data slopes.}
\label{fig-results1}
\end{figure}

Figure \ref{fig-results1}\textbf{A} shows the spikes and the intervals obtained in a 500$s$-long realization of the model, which yielded 15,164 spikes for a firing rate of approximately $30$ spikes/s (and a mean ISI of 33$ms$ with standard deviation of $4.5$). The Rescaled Range statistics and Detrended Fluctuation analyses were first applied to a shorter sequence of intervals, the first 100$s$ (3,023 spikes) of simulation. Figure \ref{fig-results1}\textbf{B} shows that the common linear regression between $\log n$ and $\log R/S(n)$ or $\log DFA(n)$ yields a $H$ value near $0.75$ in both cases, suggesting a long-range dependence of the ISI sequence. However, this is a Markovian model and we do not expect it to yield LRD, as we shall prove.

\begin{figure*}[!htbp]
\centering
\includegraphics[width=0.9\linewidth]{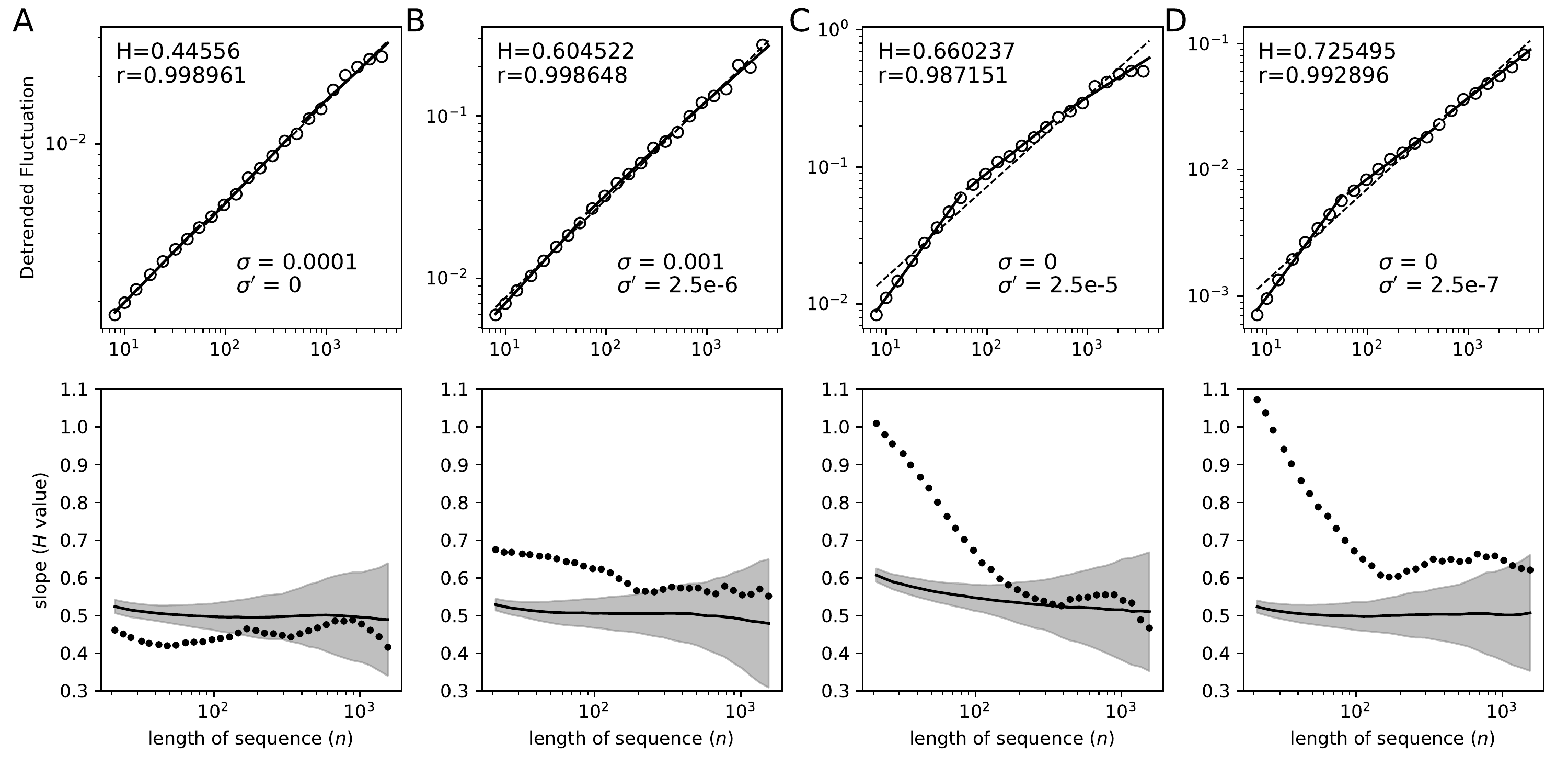}
\caption{ Hurst estimation depending on the different sources of noise. Detrended Fluctuation Analysis of simulations with noise only in the voltage equation (\textbf{A}), in both the Voltage and the Adaptation equation (\textbf{B}), and in the Adaptation equation (\textbf{C} and \textbf{D}) with two values of \(\sigma'\). Panels are as described in Figure \ref{fig-results1}\textbf{C} and \ref{fig-results1}\textbf{D} (right). }
\label{fig-results2}
\end{figure*}

\begin{figure*}[!t]
\centering
\includegraphics[width=0.7\linewidth]{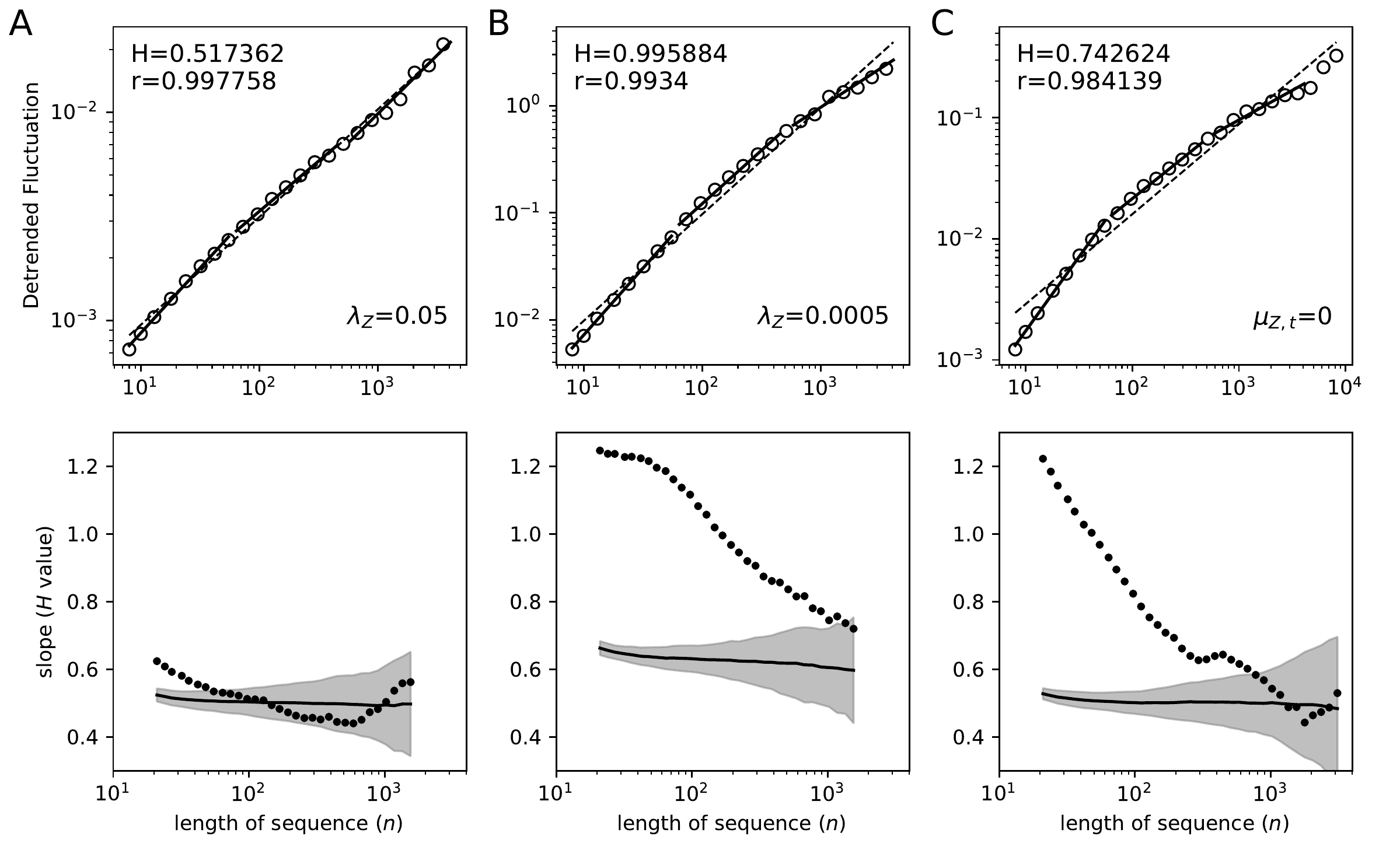}
\caption{Dependency of the LRD on adaptation parameters. \textbf{A.} Effect of a larger rate for $Z$. \textbf{B.} Effect of a smaller rate. \textbf{C.} Long-range dependence analysis in the absence of adaptation, i.e. the $Z$ variable is not affected by the occurrence of spikes.}
\label{fig-results3}
\end{figure*}

Visual inspection of the plots reveals that the slope calculated is far from being the asymptotic slope, and that the curve `bends' towards the right end. When we included the full sequence to the $R/S$ and DFA analyses (Figure~\ref{fig-results1}\textbf{C}), it is evident that the points are not following a linear relationship and the calculated slopes are lower. To characterize better the non-asymptotic nature of the slope, we repeated the fit to smaller subsets of 15 contiguous points, in a sliding window fashion. Three of such fits are shown as continuous lines in Figure \ref{fig-results1}\textbf{C} (note that in Figure~\ref{fig-results1}\textbf{C} every other point has been omitted) and in Figure~\ref{fig-results1}\textbf{D} are plotted the slopes at different positions of the moving window. From this figure, it is clear that for both $R/S$ and DFA the actual asymptotic behavior is a slope of 0.5. As the sequence length $n$ increases, the slope approaches the 0.5 value and, moreover, gets into the 2*standard deviation range calculated from surrogate data (see Section \ref{sec:stats}). Thus, only the analysis of a very large sequence of data ---probably discarding the shorter sequences in the analysis--- will reveal that what appears to be long-range dependence has only a limited time span and that the phenomena underlying it is Markovian. Nevertheless, even on smaller sequences, we see that the Hurst estimator $\hat{H}_n$ is decreasing with $n$ in the Markovian model (Figures \ref{fig-results2} and \ref{fig-results3}), while it is relatively stable in the fractional case, as we shall see later (Figure \ref{fig-results4}).
In the following, the $R/S$ analysis is no longer displayed in the plots. The reason is that we systematically observed a similar quantitative behavior between the $R/S$ and the DFA, hence it was not necessary to keep both. We chose the DFA over the $R/S$ for its better accuracy (see Figure \ref{fig:boxplot_Hurst}).

The apparent long-range dependence of the data is largely related to the stochastic nature of the adaptation. Figure \ref{fig-results2}\textbf{A} shows that the apparent LRD is lost when the noise is present only in the voltage equation ($\sigma > 0$) but not in the adaptation ($\sigma'=0$). When the noise is present in both equations, the apparent LRD is somewhat reduced for high values of $\sigma$ (Figure \ref{fig-results2}\textbf{B}). Also, Figure \ref{fig-results2}\textbf{B} shows an interesting case where visually the straight line seems to be a good fit of the $\log n$ versus $\log DFA(n)$ data (and the associated $r$ coefficient seems also good). However, the bottom plot shows that $H>0.5$ is only observed at the left side of the plot (shorter sequences) while the asymptotic value actually falls within the standard deviation for the shuffled data. On the other hand, the magnitude of the noise seems not to affect much this behavior (Figure \ref{fig-results2}\textbf{C} and \textbf{D}).

The LRD is also linked to the rate constant for slow adaptation, $\lambda_Z$. Figure \ref{fig-results3} shows that a large rate (or a small time constant $\tau_Z=1/\lambda_Z$) is associated with the loss of apparent LRD (Figure \ref{fig-results3}\textbf{A}), while a smaller value produces a LRD that is maintained for longer sequence lengths and also a higher $H$ value (Figure \ref{fig-results3}\textbf{B}). Further parameter explorations revealed that in order to observe the apparent LRD, the time constant for slow adaptation has to be at least twice the mean interval between spikes (not shown). 

Figure \ref{fig-results3}\textbf{C} explores the situation where the adaptation variable \(Z\) is no longer updated at each spike (i.e., \(\mu_{Z,t}=0\) for every \(t\)). In this case, \(Z_t\) can be understood as a correlated noise (in the form of an Ornstein-Uhlenbeck process) added to the variable \(V\). Although the adaptation effect is lost and the firing rate is increased (not shown), the apparent LRD is still present, showing that it is the correlated nature of the stochastic variable that causes this effect. This is very much in line  with what has been described for other statistics of firing in the presence of different forms of correlated noise \citep{SchwalgerEtAl2015,Schwalger}.

\subsubsection{Stationarity}

\begin{figure}[!ht]
  \centering
  \includegraphics[width=\linewidth,height=0.45\textheight]{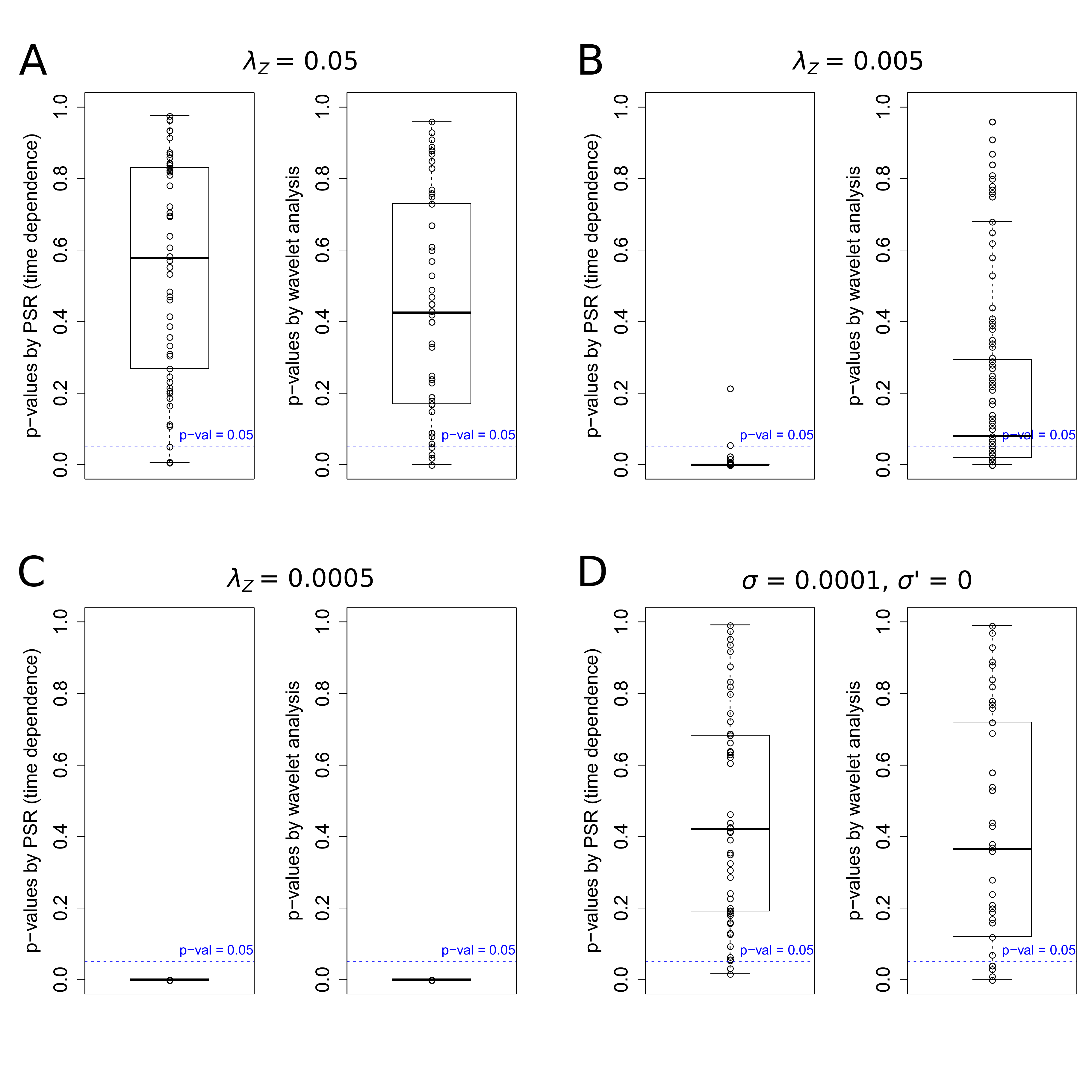}
  \caption{PSR and wavelet tests of stationarity on the PIF model with stochastic adaptation. The box displays the first quartile, the median and the third quartile, the whiskers extend to the most extreme data point which is no more than 1.5 times an interquartile away from the box. We observe here the effect of $\tau_Z = 1/\lambda_Z$ on the stationarity of the ISIs. We see in \textbf{A}, \textbf{B} and \textbf{C} that the smaller $\lambda_Z$ is, the further the ISIs are from being stationary. $\tau_Z$ can be interpreted as a relaxation time towards a stationary regime. In \textbf{D}, the absence of noise in the adaptation variable yields stationary ISIs.}
  \label{fig:PSRWVLT}
\end{figure}

\begin{figure*}[!h]
\centering
\includegraphics[width=0.8\linewidth]{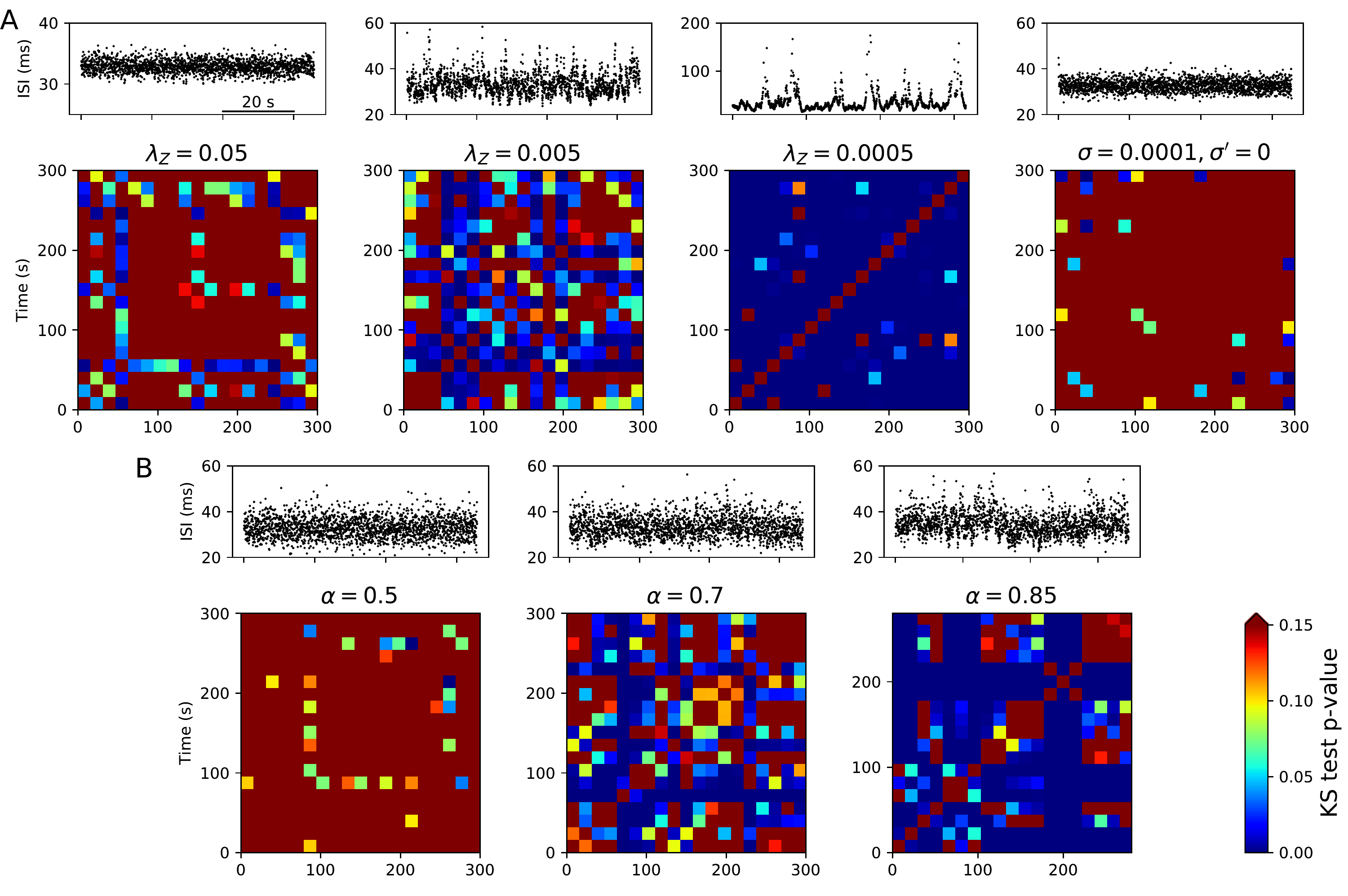}
\caption{Windowed KS test of ISI series, for the PIF model with stochastic adaptation (\textbf{A}) and the PIF model with fractional Gaussian noise (\textbf{B}). In \textbf{A}, the panel for $\lambda_Z$=0.005 analyses the same data as in Figure \ref{fig-results1}, while panels for $\lambda_Z$=0.05 and $\lambda_Z$=0.0005 use the same data as in Figure \ref{fig-results3}\textbf{A} and \textbf{B}, respectively. The panel with $\sigma$=0.0001, $\sigma'$=0 corresponds to Figure \ref{fig-results2}\textbf{A}. In \textbf{B}, three values of $\alpha$ are shown. At the top of each panel, a sample sequence of 60 s long (around 1800 spikes) is shown. The ISI sequence analyzed in the windowed KS test is of 300 s (9000 spikes), with 20 windows of 15 s. Blue colors (p-value $<0.05$) indicate that the series compared are likely to be described by different distributions.}
\label{fig:resultsKS}
\end{figure*}

We base our stationarity analysis of the spike trains on the windowed KS, PSR and wavelet tests. For the PSR and wavelet tests, we apply the methodology described in Section \ref{sec:stats}, hence in each panel of Figure~\ref{fig:PSRWVLT}, the left bar is a boxplot of 50 p-values from the PSR test computed from 50 independent spike trains generated by the same model; the right bar does the same with the wavelet test. On the other hand, Figure~\ref{fig:resultsKS} shows the results of the windowed KS test (see \ref{subsubsec:windowKS}) for a single realization of the models indicated.

In the PIF model with stochastic adaptation, a lower adaptation rate $\lambda_Z$ (longer adaptation time constant) is associated with a loss of stationarity, i.e. the data windows are no longer described by the same distribution (Figures \ref{fig:PSRWVLT}\textbf{A}-\textbf{C} and \ref{fig:resultsKS}\textbf{A}). It seems that a lower adaptation rate $\lambda_Z$ (corresponding to a larger relaxation time $1/\lambda_Z$) produces a sequence of ISIs farther from stationarity, and that $1/\lambda_Z$ not only characterizes the speed of convergence of $Z_t$ to its stationary regime, but also the speed of convergence of the law of the ISIs to their stationary law.

Observing Figures \ref{fig:PSRWVLT} and \ref{fig:resultsKS}\textbf{A}, it seems that the spike trains for larger $\lambda_Z$ are stationary while they are not for smaller $\lambda_Z$. A first explanation could be that the transient period to reach a stationary regime is longer for small $\lambda_Z$, since the characteristic time $\tau_Z = \lambda_Z^{-1}$ is larger. However we obtained the same result after removing a sufficiently large number of spikes at the beginning of the sequence. In fact we believe that even for small $\lambda_Z$, the spike train reaches a stationary regime: First, it is likely that the stationarity tests are not robust to the very large fluctuations induced by a small $\lambda_Z$ (see Figure~\ref{fig:resultsKS}\textbf{A}) and do not scale well; second, we performed multi-sample tests of stationarity for $\lambda_Z=0.0005$, which confirmed our intuition of a long transient period followed by stationarity (although we recall that such tests require to simulate many spike trains from the same model, which is not feasible biologically). 
This question has practical consequences, since non-stationary time series cannot be analysed, in general, with the same methods as stationary time series and doing so can lead to severe mistakes.

On the other hand, adding only white noise to the dynamics of $V$ produces stationary data (Figures \ref{fig:PSRWVLT}\textbf{D} and \ref{fig:resultsKS}\textbf{A}).

\subsection{PIF model with fractional Brownian noise}\label{subsec:resultsfPIF}

We decided to compare the behavior of the previous Markovian PIF model with or without adaptation (but with noise always in the adaptation variable) to a non-Markovian PIF model without adaptation mechanism, as this was proved to be irrelevant as far as LRD is concerned. Therefore we set $\gamma=0$, $\lambda_V=0$ and explored values of $\alpha$ above $0.5$. $\sigma$ and $\mu$ were adjusted in order to obtain similar mean and variance of the ISIs obtained in the previous simulations.

\subsubsection{Long-range dependence}

\begin{figure*}[!htbp]
\centering
\includegraphics[width=0.7\linewidth]{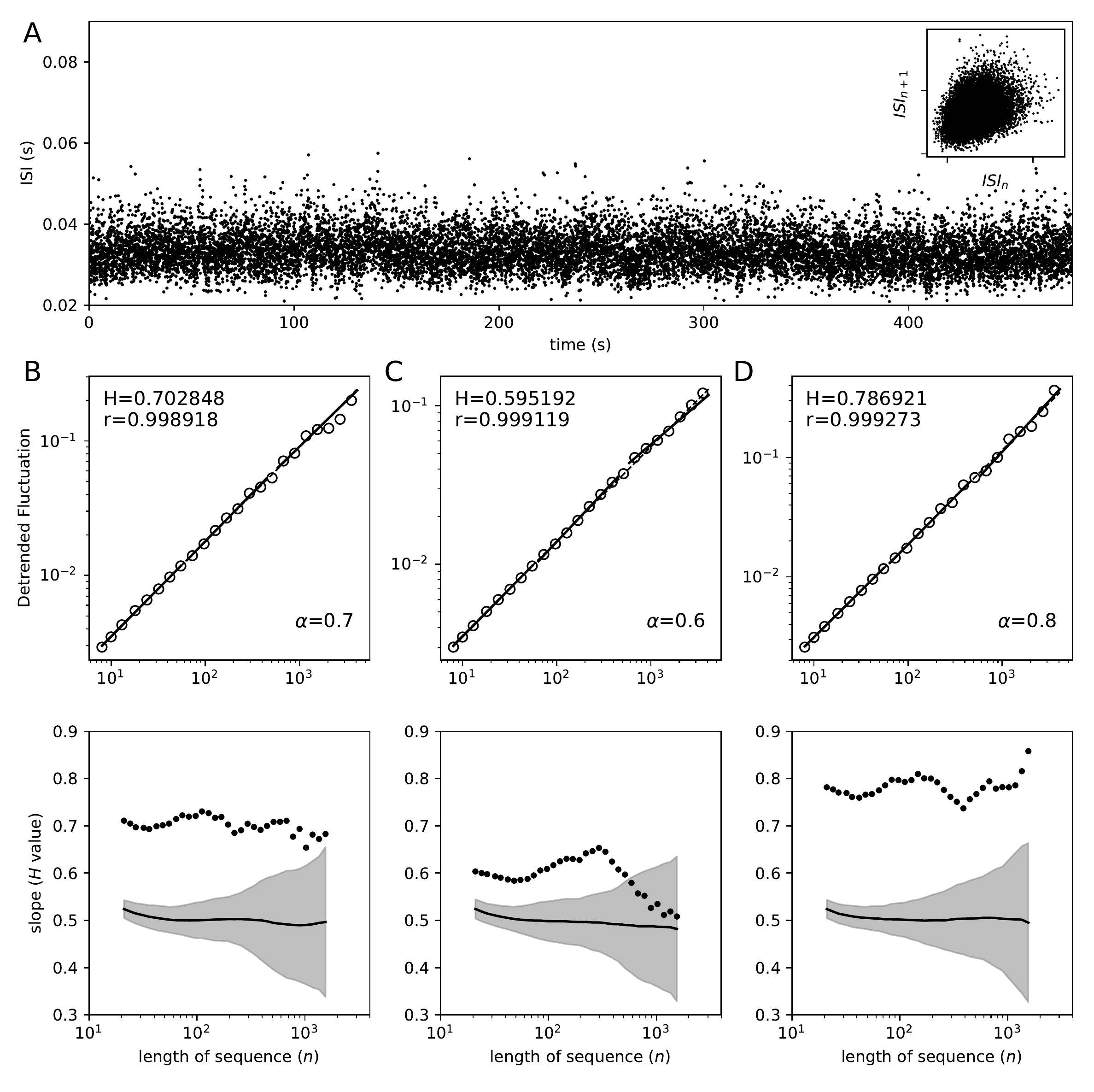}
\caption{ Long-range dependence behavior in a PIF model with fractional Gaussian noise. \textbf{A.} sequence of ISIs obtained in a simulation with equation (\ref{eq:generalLIF}) and parameters $\mu_V=0.0303$, $\lambda_V=0$, $\gamma=0$, $\sigma=0.0117$, $\alpha=0.7$. The $Z$ variable was not taken into account. Inset, $ISI_{n}/ISI_{n+1}$ plot. \textbf{B.} DFA analyses for the full sequence of 14,500 spikes. The three continuous lines that depict local slopes are overlapping a segmented line that represents the best fit for all the data points. As in Figure \ref{fig-results1}\textbf{C}, every other point has been omitted. Bottom, plot of best-fit slopes in moving windows of 15 points. The continuous line and the shadowed region are the mean and standard deviation, respectively, of the fits with surrogate data. \textbf{C} and \textbf{D.} DFA analysis of ISI sequences obtained with $\alpha=0.6$ and $\alpha=0.8$, respectively.}
\label{fig-results4}
\end{figure*}

\begin{figure}[!htbp]
\centering
\includegraphics[width=\linewidth]{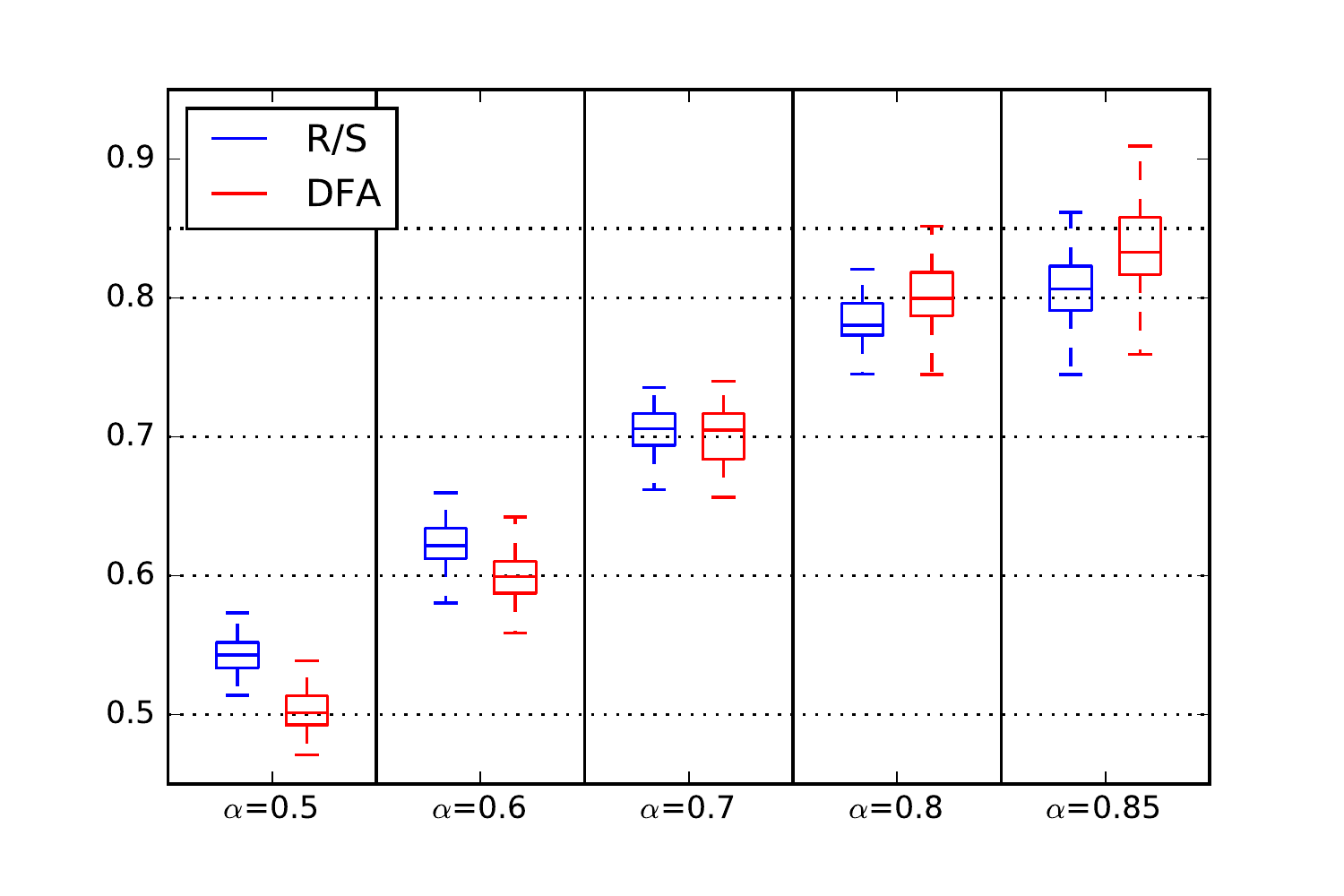}
\caption{Measure of the Hurst parameter $\hat{H}_n$ of the fractional PIF model. For each $\alpha\in\{0.5, 0.6, 0.7, 0.8, 0.85\}$, we simulated $50$ independent sequences of ISIs from a fractional PIF  with parameter $\alpha$ (and with $\mu$ and $\sigma$ chosen so that the ISIs $X_n$ have the following moments $\EE[X_n]\approx 32.9 \text{ms}$ and $\text{Var}[X_n]\approx 20$). The Hurst parameter was estimated by the $R/S$ (blue plot) and DFA (red plot) methods for each simulation, and for each underlying $\alpha$ parameter, the result has been aggregated in a boxplot. We see that the estimated Hurst parameter of the ISIs is very close to the value of the scaling parameter of the fBm used in the simulations. The DFA method seems to perform better.}
\label{fig:boxplot_Hurst}
\end{figure}

Figure \ref{fig-results4} shows that adding a fractional Gaussian noise indeed produces a long-term dependence in the series of ISIs, as evidenced by both Rescaled Range statistics and Detrended Fluctuation Analysis. In contrast to the PIF model with stochastic adaptation, however, the high slope in the $\log n$ versus $\log R/S(n)$ or $\log DFA(n)$ plots is maintained and does not decay as $n$ increases. In other words, the $\hat{H}_n$ value obtained by these analyses appears to be rapidly close to its true asymptotic value. This behavior is observed at different values of $\alpha$ (Figure \ref{fig-results4}). The PIF model with fractional Brownian noise, however, shows a weaker correlation of consecutive intervals (Figure \ref{fig-results4}\textbf{A}, inset, r=0.34) than the PIF model with noisy adaptation (compare to inset of Figure \ref{fig-results1}\textbf{A}, r=0.85).\\
Furthermore, we see in Figure \ref{fig:boxplot_Hurst} that estimated Hurst parameter $\hat{H}_n$ is very close to the input value $\alpha$. Hence we can safely assert that $\hat{H}_n$ converges to $\alpha$.

\subsubsection{Stationarity}

Results concerning stationarity of this model are shown in Figure \ref{fig:resultsKS}\textbf{B} for the KS test. We conclude that the ISIs are stationary when $\alpha=0.5$ (the PSR and wavelet tests yielded the same conclusion). This agrees with the theoretical result in this simple framework. The conclusion from the cases $\alpha=0.7$ and $\alpha=0.85$ is less straightforward in view of the PSR and wavelet tests, but we performed additional tests (all not shown) which suggest stationarity as well. Even more than for the Markovian model though, proving stationarity seems mathematically challenging.

\subsection{Other models}\label{subsec:otherPIFS}

\subsubsection{Leaky Integrate-and-Fire models}

The leaky Integrate-and-Fire model corresponds to $\lambda_V>0$ in Equation (\ref{eq:generalLIF}), instead of $\lambda_V=0$ for the PIF. We draw the same conclusions on the LRD property for the LIF (not shown).

\subsubsection{Higher dimensional Integrate-and-Fire models with Brownian noise}\label{subsec:HD_IF}

\begin{figure}[!h]
  \centering
  \includegraphics[width=0.9\linewidth]{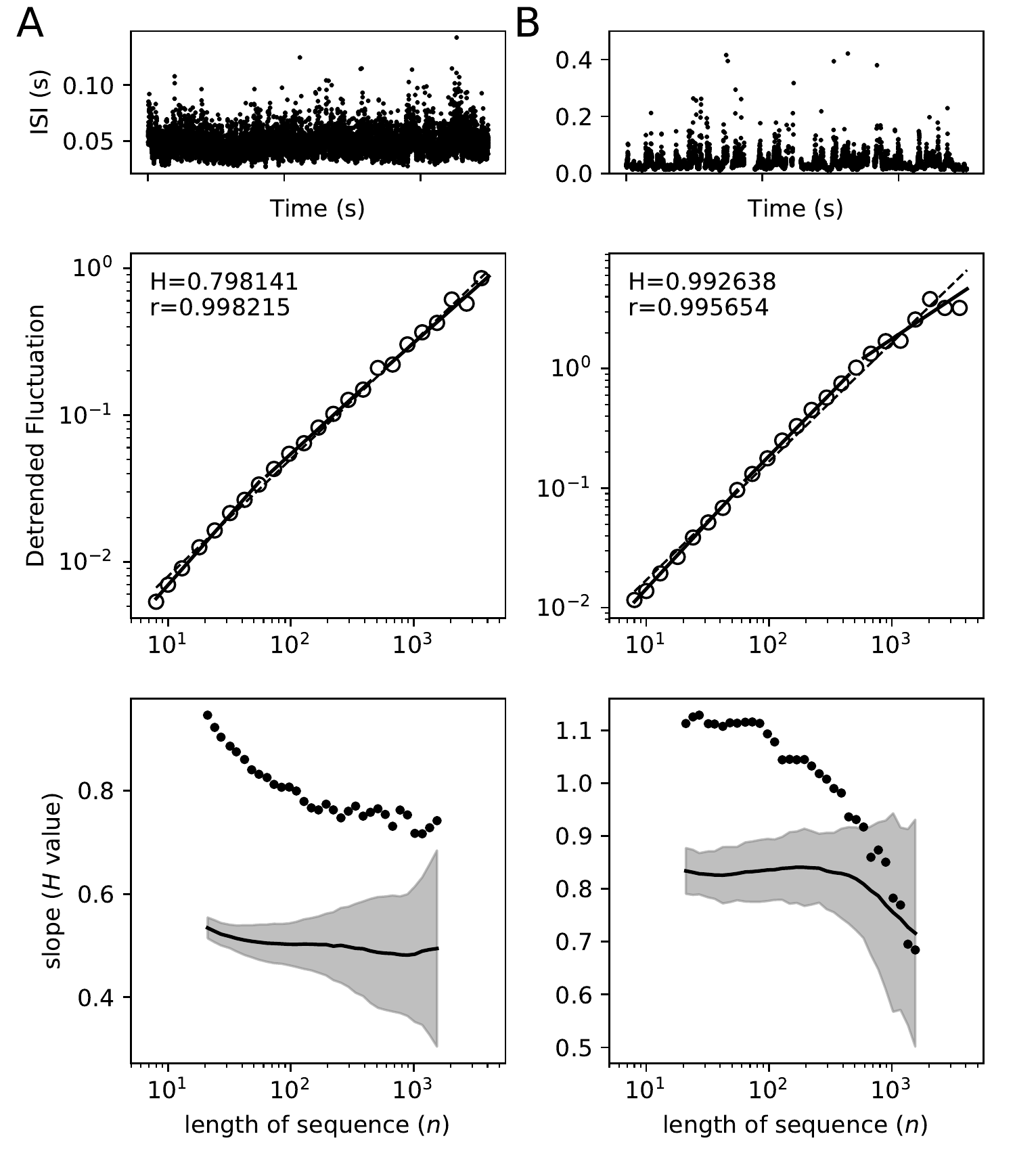}
  \caption{Long-range dependence in the PIF model with multidimensional noise. In \textbf{A}, a PIF with a single time constant ($\tau=5000$ms) is simulated and the Hurst parameter decreases to $0.5$, even with a large time constant. This is coherent with the results of Section~\ref{subsec:PIF_LRD}. On the other hand in \textbf{B}, a PIF with three-dimensional noise whose time constants are $200ms$, $1000ms$ and $5000ms$ is simulated and we observe a slower decay of the Hurst estimation. But in both cases, the slope is decreasing, unlike the fractional PIF model (Fig.~\ref{fig-results4}).} 
  \label{fig:Multidim_PIF}
\end{figure}

Following the idea in \citep{SchwalgerEtAl2015}, we simulated a PIF with three noisy adaptation variables whose time constants are $200ms$, $1000ms$ and $5000ms$ (Figure \ref{fig:Multidim_PIF}\textbf{A}). The aim is to get a better approximation of long-range dependence with a Markovian model. We observe in Figure \ref{fig:Multidim_PIF}\textbf{A} that the Hurst estimation decays more slowly in this new model, indicating that it can be a good approximation of a LRD sequence when the length is not too large. Yet it still seems to converge to $0.5$, which means that it is still not LRD. To emphasize the slower convergence of the LRD estimator in the multidimensional model, we compared it to the previous PIF model (from Subsection~\ref{subsec:PIF}) with a large time constant of $5000ms$ (Figure \ref{fig:Multidim_PIF}\textbf{B}).

We observe (particularly in Figure \ref{fig:Multidim_PIF}\textbf{B}) that the mean curve of the surrogate data can be higher than $0.5$ for the DFA, although it decreases (we checked that it indeed decreases to $0.5$ for longer spike trains). This is a drawback of the DFA, which is not robust to data with very large mean and/or variance. We confirmed this either by artificially removing the largest ISIs (hence reducing drastically the variance), or by simulating sequences of \emph{i.i.d.} positive random variables following a Pareto law (with scale $1$ and shape parameter between $1$ and $2$). In that situation, a strong bias appears, unlike the $R/S$ which still returns values close to $0.5$
(not shown).

\subsubsection{PIF model with Brownian and fractional Brownian noise}

Considering the approach of some authors \citep{SchwalgerEtAl2015} to add high-dimensional noise in the adaptation variable (see also some heuristics in Section \ref{subsec:heuristics}) and the power law behavior of the adaptation often observed \citep{FairhallEtAl01}, one is tempted to consider the following modification of our model (\ref{eq:generalLIF}), where the scaling parameter of the noise in the voltage is $0.5$ and is $\alpha>0.5$ in the adaptation variable:
\begin{align}\label{eq:fLIF_adaptation}
\begin{split}
\dd V_t &= \left(\mu - \lambda_V V_t + \gamma Z_t\right)\dd t + \sigma \dd B^{1/2}_t \\
\dd Z_t &= \left(- \lambda_Z Z_t\right)\dd t + \sigma' \dd \tilde{B}^\alpha_t  
\end{split}
\end{align}
This model may allow for more complex behaviors, e.g. observations from simulations on the previous model display several firing regimes:
\begin{itemize}[topsep=3pt,itemsep=0pt,partopsep=0pt,parsep=1pt]
	\item if $\sigma = 0$ and for $\lambda_Z>0, \lambda_V=0,  \mu= -10\gamma$, the estimated Hurst parameter of the ISIs, $\hat{H}_n$, remains close to the scaling parameter $\alpha$ of the model.
	\item if $\sigma>0$ and $\sigma' \ll \sigma$, the ISIs are almost independent and $\hat{H}_n$ is close to $0.5$.
	\item if $\sigma \ll \sigma'$ and $\mu\ll 1$, then similarly to the biological data, $\hat{H}_n$ increases with $n$ towards the value $\alpha$ and the ISI histogram seems to deviate from the inverse Gaussian distribution (figures not shown).
\end{itemize}
Furthermore, it would be interesting to see in future works if this model shares the multiple time scale adaptation observed by \citep{FairhallEtAl01} and modelled by \citep{LundstromEtAl,TekaEtAl14} using fractional differentiation or cascade processes as in \citep{DrewAbbott,PozzoriniEtAl}.

\section{Discussion}\label{sec:discussion}

In this paper, we have studied two approaches to model the \emph{long-range} temporal correlations observed in the spike trains of certain neurons. In a first approach, the introduction of a weakly correlated input noise with a finite number of timescales into a linear Integrate-and-Fire model can produce quite large time dependencies and be a good approximation of a power law dynamics; however this is not genuine LRD nor power law behavior, as it is shown when a sufficiently large sequence is analyzed. Besides, we have shown that using multiple large time constants usually yields non-stationarities. A second approach is also of the Integrate-and-Fire type, with a stochastic input called fractional Brownian motion. To the best of our knowledge, this is the first time this stochastic process is used in an IF model, and we showed that it is very well suited to produce genuine long-range dependent spike trains. Besides, this type of stationary Gaussian noise emerges naturally as a scaling limit of discrete noises (see Sections \ref{subsec:other_models} and \ref{subsec:heuristics}) which can originate either from the fractal behavior of ion channels or from the cumulated inputs of the neuronal network.

To measure the long-range dependence of spike trains, we followed a well-established procedure \citep{Taqqu}. For previous examples in neuroscience, see for instance the DFA analysis of ISIs in \citep{BhattacharyaEtAl} or the $R/S$ analysis for ion channels in \citep{CorrelChannels}.
 In the literature to date, we have identified several types of IF models aimed at producing correlated spike trains: those with colored noise input (i.e. Markov noise) \citep{BrunelSergi,MiddletonEtAl2003,Lindner2004,SchwalgerEtAl08,Schwalger} and more recently PIF with high-dimensional Ornstein-Uhlenbeck noise \citep{SchwalgerEtAl2015}; non-renewal point processes input\footnote{a limitation of the point process approach is that it is far from the biological reality.} \citep{Teich92,BairEtAl94,TeichEtAl97,LowenEtAl97,LowenEtAl01,Jackson}; and models with $1/f$ noise input \citep{SobieEtAl}.
The first conclusion of our study is that the ISIs generated from \emph{Markovian} integrate-and-fire models do not have LRD \emph{stricto sensu}, and produce instead ISIs whose correlations are exponentially decaying. From this perspective, we must however point out \citep{SchwalgerEtAl2015} whose precise goal was to replicate power-law decay of the correlations. While this goal is achieved on a reasonable band of frequency (see the power spectrum of their simulated spike trains), we have shown that such models still do not produce LRD. However, as seen in all Figures \ref{fig-results1} to \ref{fig-results3}, these Markovian integrate-and-fire models (whether perfect of leaky) can replicate a LRD effect for sequences of spikes with a given length (see Figure \ref{fig-results1}), if the adaptation variable is noisy and its time constant $1/\lambda_Z$ is large enough. Nonetheless, plots of the estimated Hurst parameter as a function of the sequence length are always decreasing. This contrasts with fractional integrate-and-fire models, for which this function appears constant at a value $\hat{H}$ (see Figure \ref{fig-results4}). This provides a simple criterion to discriminate between Markovian and fractional IF models. Moreover, we see in Figure \ref{fig:boxplot_Hurst} that $\hat{H}$, the estimated Hurst index of the spike trains, is exactly the scaling parameter $\alpha$ of the fractional Brownian motion injected in the model.

We also presented and compared the effectiveness of several stationarity tests suited to time series analysis. The methodology for testing stationarity we propose seems relatively new to the neuroscience literature. Stationarity is often believed to hold for ISIs \citep{SchwalgerEtAl2015}, yet it produced surprising results since we observed that sequences of ISIs can look non-stationary (Figures \ref{fig:PSRWVLT} and \ref{fig:resultsKS}), even when generated from a simple IF model with Ornstein-Uhlenbeck noise. However, we believe that a stationary regime exists for such models. 
This stationarity property has important consequences: if a sequence of ISIs has a stationary regime and its correlations decay exponentially fast, then the estimated Hurst of the $R/S$ statistic must be $0.5$. Altogether the present discussion on stationarity leaves several questions unanswered and should be the purpose of future work.

A very interesting and important problem that we may also be the content of future work is \emph{calibration}. Consider the following situation: given an observed spike train with measured Hurst parameter $\hat{H}>0.5$, we want to calibrate either the parameters $\mu_V, \sigma$ and $\alpha$ of a fractional PIF or the parameters $\mu_V, \gamma, \sigma, \lambda_Z, \sigma'$ of a Markovian ($\alpha=0.5$) PIF with an adaptation variable. In the first case, it results from Figure \ref{fig:boxplot_Hurst} that we must choose $\alpha=\hat{H}$. We only have two more parameters to fix, and the mean of the ISIs is given by $\frac{1}{\mu_V}$ (assuming implicitly that the threshold is $1$). We can then try to compute $\sigma$ from the variance of the ISIs and $\hat{H}$. On the other hand, we have seen from Figure \ref{fig-results3} that the $\hat{H}$ value can be replicated by adjusting $\lambda_V$: a larger $\lambda_V$ yields smaller $\hat{H}$ parameter, but also impacts the first two moments of the ISIs. Hence it may be easier to fix first the scaling parameter of the noise, rather than having additional parameters just to replicate the correlations of the ISIs. Then we can focus on additional properties that adaptation can bring to integrate-and-fire models.

\subsection{Other classes of models with fractal/LRD behavior}\label{subsec:other_models}

Despite the numerous articles emphasizing the presence of fractal and/or long-range dependence of the spiking activity of some neurons (see Introduction), we merely identified two streams of papers proposing a model reflecting these characteristics. 
In \citep{Jackson} (see references therein from related previous works from the 90's, including  in particular \citep{LowenEtAl97} and coworkers), an integrate-and-fire model is used in conjunction with various point processes modelling a random input into the neuron. If the point process is a renewal process, then it may produce long-range dependence only if it has infinite variance \citep[Theorem 2]{Jackson}. Infinite variance models can get far from biological observations, thus a more sophisticated point process, the fractional-Gaussian-noise-driven Poisson process, is used in \citep{Jackson}. This process is a doubly stochastic Poisson process, whose (stochastic) rate function is a nonlinear function of a fractional Gaussian noise. Each jump corresponds to a spike in a presynaptic neuron, and when injected in an IF model, it is successful in producing spike trains with long-range dependence (as measured with the Fano factor). However, the use of such process seems less mathematically tractable than our approach with a fractional noise. In fact, the fBm is itself the scaling limit of discrete processes \citep{Taqqu75,Sottinen,HammondSheffield}, is statistically self-similar and with stationary increments, which makes it a natural candidate as input noise.

The second approach to model LRD is an Integrate-and-Fire model with $1/f$ noise proposed by \cite{SobieEtAl}, strongly related to our model. The link between fractional Brownian motion and $1/f$ noise is explained in \citep{AbryEtAl95}, although there is no definition of $1/f$ noise as clear and universally accepted as the definition of fBm can be. Besides, an advantage of using fBm is that it can be exactly simulated, which ensures that all frequencies are present in its spectrum and that LRD holds, while a simulated $1/f$ noise is an approximate $1/f$ noise with limited bandwidth. Nevertheless, the approach of \cite{SobieEtAl} is complementary to ours since this study focuses on the dispersion of spike trains in time windows $[0,t]$ for various times $t$ (as measured by the Fano factor).

\subsection{Heuristics on long-range dependence and fractional Brownian motion}\label{subsec:heuristics}

In classical Markovian models (e.g. PIF model with multidimensional Ornstein-Uhlenbeck noise, $\alpha=0.5$), the correlation between interspike $i$ and $i+n$ decays exponentially in $n$, even though having high-dimensional OU process is intended to produce large time constants. We assert that the ISIs of this model are mixing, i.e. that $\sup_{A,B} |\PP\left(X_i\in A, X_{i+n}\in B\right) -\PP(X_i\in A) \PP(X_{i+n}\in B)|\leq \phi(n)$, for some $\phi$ such that $\sum_n \phi(n)<\infty$. We also believe, based on mathematical arguments and some numerical evidence (see Section~\ref{sec:results}), that such model produces ISIs which converge to a stationary regime.

From \citep{Doukhan}, Chapter 1.5, it is known that any stationary and mixing sequence satisfies an invariance principle. This is enough to apply Theorem 4 of \cite{Mandelbrot75}, which gives the convergence of $N^{-1/2} R/S(N)$ to a non-trivial random variable. Therefore we conjecture the following result that we plan to prove in a separate work, that is:
\begin{quote}
\emph{If $\alpha=0.5$ , the sequence of interspike intervals generated by the PIF/LIF model (\ref{eq:generalLIF}) has a stationary regime, and $N^{-\frac{1}{2}} R/S(N)$ converges to a non-degenerate random variable (i.e. $\hat{H}_N \rightarrow 0.5$).}
\end{quote}

Our second heuristics is about the approximation of the fractional Brownian motion by a sequence of $n$-dimensional Ornstein-Uhlenbeck processes, as $n$ increases. In \citep{SchwalgerEtAl2015}, the general idea is that the covariance of a general Gaussian process can be approximated by an Ornstein-Uhlenbeck with sufficiently many components. In \citep{CarmonaCoutinMontseny}, it is proven that the fBm is indeed an infinite-dimensional Ornstein-Uhlenbeck process. Therefore, we can consider our model with fractional noise as a natural limit to the model proposed in \citep{SchwalgerEtAl2015}. Although this is not the only possible limit in their approach, the fBm is the most sensible choice to obtain long-range dependence.

\begin{acknowledgements}
	Part of this work was carried out while A.R. was
	a postdoc at Inria Sophia-Antipolis and at Ecole Polytechnique (the support from ERC 321111 Rofirm is gratefully acknowledged). A.R. and E.T. acknowledge the support from the ECOS-Sud Program Chili-France C15E05 and from the European Union's Horizon 2020 Framework Program for Research
	and Innovation under Grant Agreement No. 720270 (Human Brain Project SGA1). P.O acknowledges the support from the Advanced Center for Electrical and Electronic Engineering (Basal Funding FB0008, Conicyt) and the project P09-022-F from the Millennium Scientific Initiative of the Chilean Ministry of Economy, Development, and Tourism.
	
	\noindent We thank the reviewers for their remarks which helped to improve
	significantly the quality of this paper.
%
%
\end{acknowledgements}


\end{document}